# Tooth-shaped plasmonic waveguide filters with nanometeric sizes

Xian-Shi LIN and Xu-Guang HUANG[*]

*Laboratory of Photonic Information Technology, South China Normal University, Guangzhou, 510006, China*

[*]*Corresponding author: huangxg@scnu.edu.cn*

A novel nanometeric plasmonic filter in a tooth-shaped Metal-Insulator-Metal waveguide is proposed and demonstrated numerically. An analytic model based on the scattering matrix method is given. The result reveals that the single tooth-shaped filter has a wavelength filtering characteristic and an ultra-compact size in the length of a few hundred nanometers, compared to grating-like SPPs filters. Both analytic and simulation results show that the wavelength of the trough of the transmission has linear and nonlinear relationships with the tooth depth and the tooth width, respectively. The waveguide filter could be utilized to develop ultra-compact photonic filters for high integration. © 2008 Optical Society of America

Surface Plasmons are waves that propagate along a metal-dielectric interface with an exponentially decaying field in the both sides [1,2]. The unique properties of Surface Plasmon Plolaritons (SPPs) have shown the potential to overcome the diffraction limit in conventional optics, which could be utilized to achieve nano-scale photonic devices for high integration. Several different Metal-Insulator-Metal (MIM) waveguide structures based on SPPs have been numerically and/or experimentally demonstrated such as U-shaped waveguides [3], splitters [4], Y-shaped combiner [5], Multimode-Interferometers [6], couplers [7,8], M-Z interferometers [9,10], Bragg mirrors [11], and photonic bandgap structures [12]. To achieve wavelength filtering characteristics, SPPs Bragg reflectors and nanocavities have been proposed. They include the metal heterostructures constructed with several periodic slots vertically along a MIM waveguide [13,14], the Bragg grating fabricated by periodic modulating the thickness of thin metal stripes embedded in a insulator [15] and the periodic structure formed by changing alternately two kinds of the insulators with the same width [16] or different widths [17,18]. Lately, a high-order plasmonic Bragg reflector with a periodic modulation of the core index of the insulators [19], and a structure with periodic variation of the width of the insulator in MIM waveguide [20] have been proposed. Most of the structures mentioned above, however, have the period number of N > 9 with the total lengths over 4μm and beyond subwavelength-scale, which result in relatively high insertion loss of several dB. In this letter, a nanoscale SPPs filter based on the MIM waveguide consisting of single rectangular tooth is proposed. The SPPs distributions and propagations are characterized by the method of Finite-Difference Time-Domain (FDTD), with perfect-matching-layer absorbing boundary conditions. An analytic model based on the scattering matrix method is derived to explain the filtering mechanism of the structure.

To begin with the dispersion relation of the fundamental TM mode in a MIM waveguide (shown in the inset of Fig. 1) is given by [15,21]:

$$\varepsilon_{in} k_{z2} + \varepsilon_m k_{z1} \coth(-\frac{ik_{z1}}{2}w) = 0, \quad (1)$$

with $k_{z1}$ and $k_{z2}$ defined by momentum conservations:

$$k_{z1}^2 = \varepsilon_{in} k_0^2 - \beta^2, \quad k_{z2}^2 = \varepsilon_m k_0^2 - \beta^2. \quad (2)$$

Where $\varepsilon_{in}$ and $\varepsilon_m$ are respectively dielectric constants of the insulator and the metal, $k_0 = 2\pi/\lambda_0$ is the free-space wave vector. The propagation constant $\beta$ is represented as the effective index $n_{eff} = \beta/k_0$ of the waveguide for SPPs. The real part of $n_{eff}$ of the slit waveguide as a function of the slit width at different wavelengths is shown in Fig. 1. It should be noted that the dependence of $n_{eff}$ on waveguide width is also suitable to the small part or region of the tooth waveguide with the tooth width of $w_t$ shown in Fig. 2. The imaginary part of $n_{eff}$ is referred to the propagation length which is defined as the length over which the power carried by the wave decays to 1/e of its initial value: $L_{spps} = \lambda_0/[4\pi \cdot \mathrm{Im}(n_{eff})]$. In the calculation above and the following simulations, the insulator in all of the structures is assumed to be air ($\varepsilon_{in} = 1$), and the frequency-dependent complex relative permittivity of silver is characterized by Drude model: $\varepsilon_m(\omega) = \varepsilon_\infty - \omega_p^2 / \omega(\omega + i\gamma)$. Here $\omega_p = 1.38 \times 10^{16}\, Hz$ is the bulk plasma frequency, which represents the natural frequency of the oscillations of free conduction electrons. $\gamma = 2.73 \times 10^{13}\, Hz$ is the damping frequency of the oscillations, $\omega$ is the angular frequency of the incident electromagnetic radiation, $\varepsilon_\infty$ stands for the dielectric constant at infinite angular frequency with the value of 3.7 [20].

The tooth-shaped waveguide filter is shown in Fig. 2. In the following FDTD simulations, the grid sizes in the *x* and *z* directions are chosen to be 5nm×5nm. The fundamental TM mode of the plasmonic waveguide is excited by a dipole source. Two power monitors are respectively set at the points of *P* and *Q* to detect the incident and transmission fields for calculating the incident power of $P_{in}$ and the transmitted power of $P_{out}$. The transmittance is defined to be $T=P_{out}/P_{in}$. The length of *L* is fixed to be 300nm, while the tooth width and depth are respectively $w_t$=50nm and *d*=100nm. As shown in Fig. 3(a), the tooth-shaped waveguide is of a filtering function: A trough occurs at the free space wavelength nearly 784nm with the transmittance of ~0%. The maximum transmittance at the wavelengths longer than 1700nm is over 90%. The contour profiles of the field distributions around the tooth-shaped area at different wavelengths are shown in Figs. 3(b)-3(d). The filtering structure is distinguished from the Bragg reflectors based on periodical heterostructure.

The phenomenon above can be physically explained in the scattering matrix theory [23] as follows:

$$\begin{pmatrix} E_1^{out} \\ E_2^{out} \\ E_3^{out} \end{pmatrix} = S \cdot \begin{pmatrix} E_1^{in} \\ E_2^{in} \\ E_3^{in} \end{pmatrix}, \qquad (3)$$

where $S = \begin{bmatrix} r_1 & t_1 & s_3 \\ t_1 & r_1 & s_3 \\ s_1 & s_1 & r_3 \end{bmatrix}$, $r_i$, $t_i$ and $s_i$ (*i*=1,2,3) are respectively the reflection, transmission and splitting coefficients of a incident beam from Port *i* (*i*=1,2,3), caused by the structure. $E_i^{in}$ and $E_i^{out}$ stand for the fields of incident and output beams at Port *i*, respectively. Using the fact that $|S|=1$, one can obtain:

$$r_1^2 r_3 + 2t_1 s_1 s_3 - 2r_1 s_1 s_3 - t_1^2 r_3 = 1. \qquad (4)$$

For the case of $E_2^{in} = 0$, one has

$$E_2^{out} = t_1 E_1^{in} + s_3 E_3^{in}, \tag{5}$$

in which $E_3^{in}$ is given as follows:

$$E_3^{in} = s_1 E_1^{in} \exp(i\phi(\lambda))(1 + r_3 \exp(i\phi(\lambda)) + r_3^2 \exp(2i\phi(\lambda)) + ...) = \frac{s_1 E_1^{in}}{1 - r_3 \exp(i\phi(\lambda))} \exp(i\phi(\lambda)), \tag{6}$$

where the phase delay $\phi(\lambda) = \frac{4\pi}{\lambda} n_{eff} \cdot d + \Delta\varphi(\lambda)$, and $\Delta\varphi(\lambda)$ is the phase-shift caused by the reflection on the air-silver surface. Combined Eq. (5) and Eq. (6), the output field at Port 2 is derived as:

$$E_2^{out} = t_1 E_1^{in} + \frac{s_1 s_3 E_1^{in}}{1 - r_3 \exp(i\phi(\lambda))} \exp(i\phi(\lambda)), \tag{7}$$

Therefore, the transmittance $T$ from Port 1 to Port 2 is given by:

$$T = \left| \frac{E_2^{out}}{E_1^{in}} \right|^2 = \left| t_1 + \frac{s_1 s_3}{1 - r_3 \exp(i\phi(\lambda))} \exp(i\phi(\lambda)) \right|^2. \tag{8}$$

It can be seen from Eq. (8) that, if the phase satisfies $\phi(\lambda) = (2m+1)\pi$ ($m=0,1,2…$), the two terms inside the absolute value sign on the right of the equation will cancel each other (as it can be seen in Fig. 3(c)), so that the transmittance $T$ will become minimum. Therefore, the wavelength $\lambda_m$ of the trough of transmission is determined as follows:

$$\lambda_m = \frac{4 \cdot n_{eff} \cdot d}{(2m+1) - \frac{\Delta\varphi(\lambda)}{\pi}}. \tag{9}$$

It can be seen that the wavelength $\lambda_m$ is linear to the tooth depth $d$, and depends on tooth width $w_t$, through the somewhat inverse-proportion-like relationship between $n_{eff}$ and $w_t$ shown in Fig. 1.

Figure 4(a) shows the transmission spectra of the waveguide filters with various tooth widths of $w_t$. The maximum transmittance can reach 97%. Figure 4(b) shows the wavelength of the trough vs. the tooth width of $w_t$. The primary trough of the transmission moves very significantly to short wavelength (blue-shift) with the increase of $w_t$ for $w_t<20$nm. The shift rate rapidly becomes small after $w_t>20$nm, and tends to be saturated when $w_t>200$nm. As revealed in the Eq. (9), the above relationship between the trough position and $w_t$ mainly results from the contribution of the inverse-proportion-like dependence of $n_{eff}$ on $w_t$. The change rate of $\Delta n_{eff}/\Delta w_t$ within the tooth width of 20nm is much higher than that of $\Delta n_{eff}/\Delta w_t$ after $w_t>20$nm, as shown in Fig. 1 and Ref.13, and becomes finally saturated after $w_t>200$nm. Obviously, tooth width $w_t$ should be chosen within the range of 20-200nm to avoid the critical behavior and the difficulty in fabrication process.

Figure 5(a) shows the transmission spectra of the filters with different tooth depths of $d$. It is found that the wavelength of the trough shifts to long wavelength with the increasing of $d$. Figure 5(b) reveals that the wavelength of the trough has a linear relationship with the tooth depth, as our expectation in Eq. (9). Therefore, one can realize the filter function in various required wavelength with high performance, by changing the width or/and the depth of the tooth. For example, to obtain a filter with a trough at the wavelength of 1550nm, the structural parameters of $w_t=w=50$nm and $d=237.5$nm can be chosen.

In summary, a novel plasmonic waveguide filter constructed with a MIM structure engraved single rectangular tooth is investigated. The filter is of an ultra-compact size with a few hundreds of nanometers in length, and then low insertion loss. Moreover, it is promised to reduce the difficulties in fabrication, comparing with previous grating-like heterostructures with a few micrometers in length. Our results suggest that the new structure could be utilized to develop plasmonic wavelength filters on flat metallic surfaces with extreme high integration for planar nanometeric photonic circuits.

The authors acknowledge the financial support from the Natural Science Foundation of Guangdong Province, China (Grant No. 07117866).

**Fig. 1**

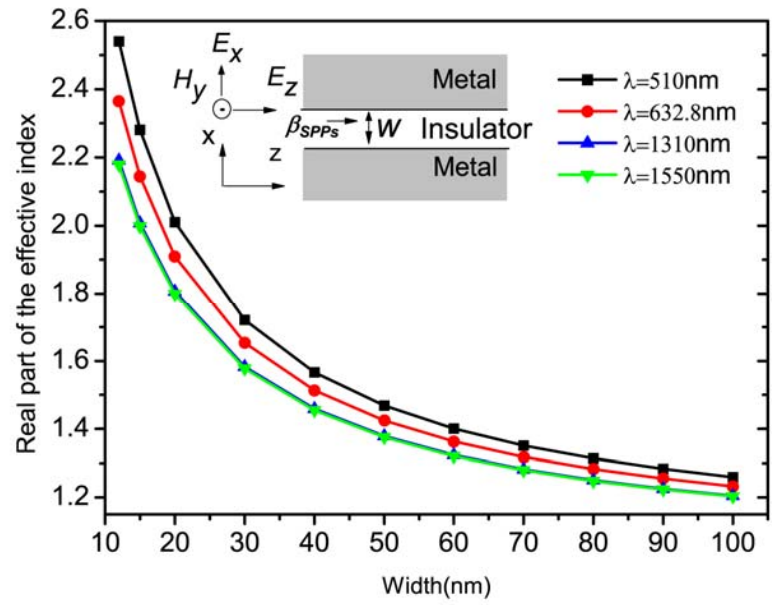

**Fig. 2**

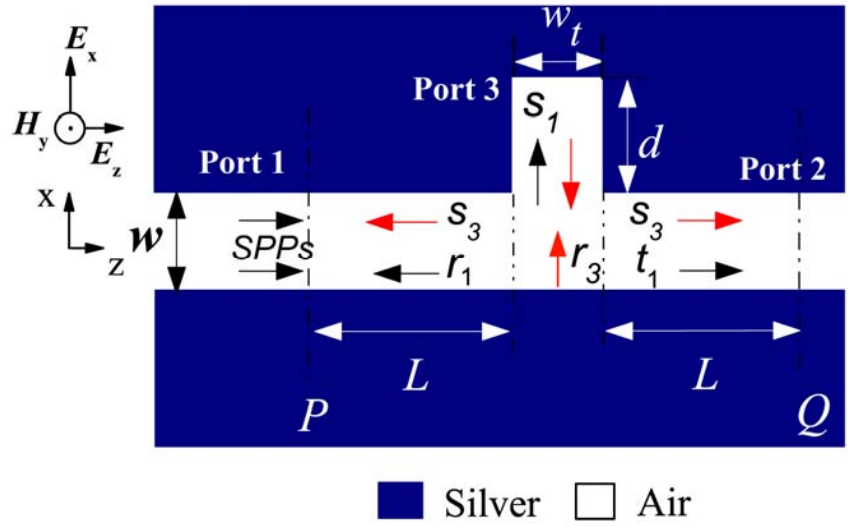

**Fig. 3**

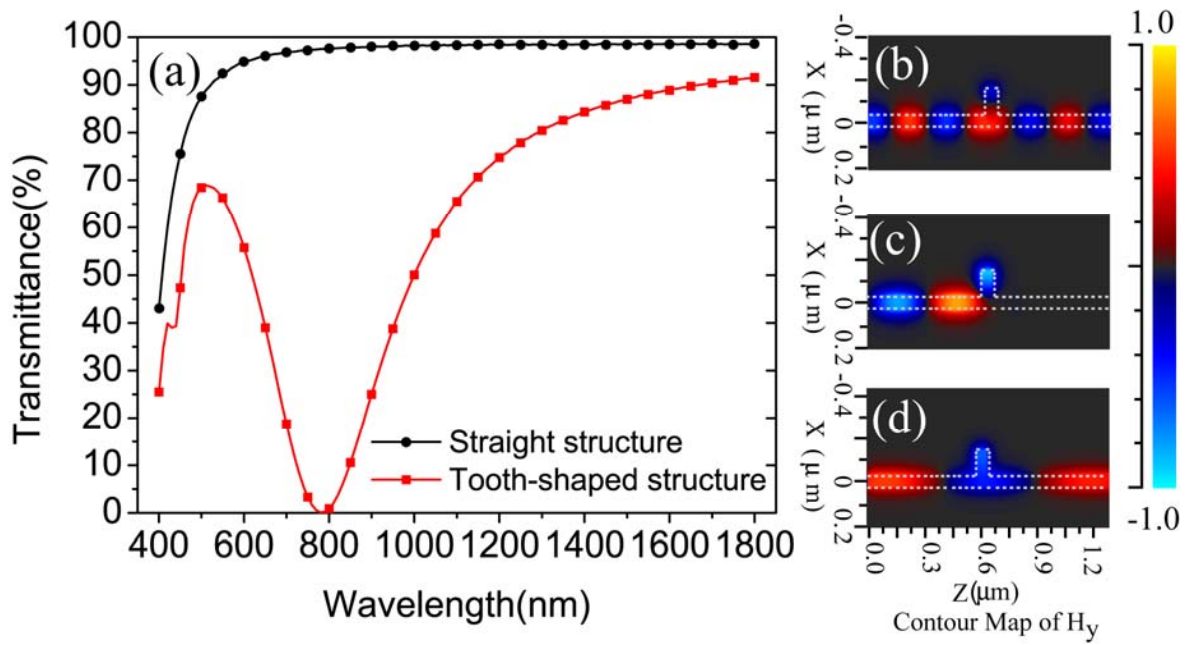

**Fig. 4**

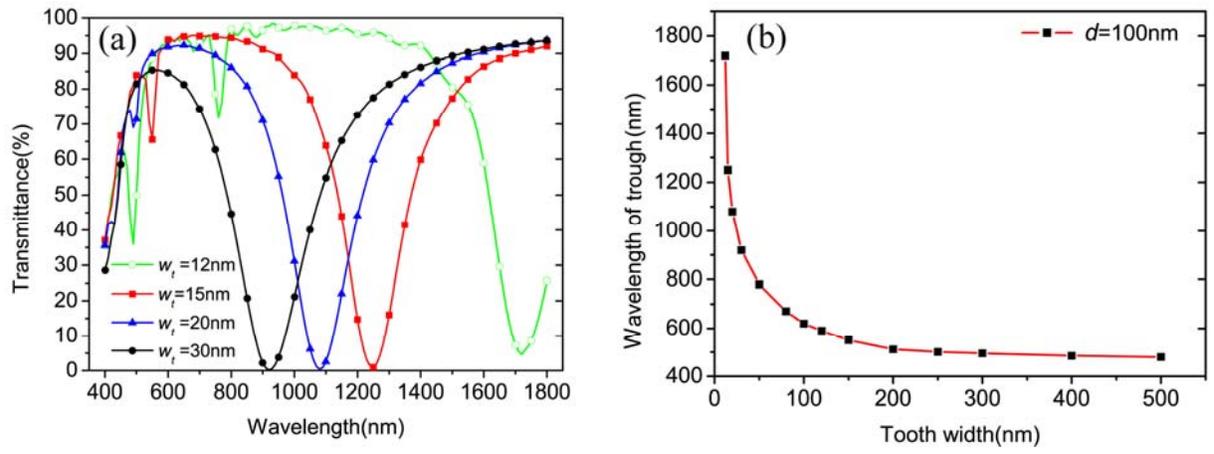

**Fig. 5**

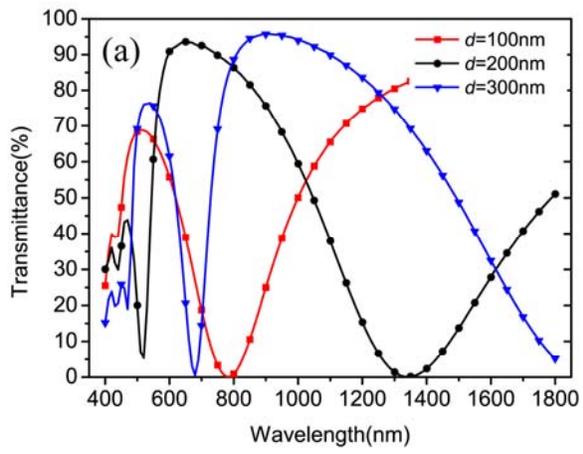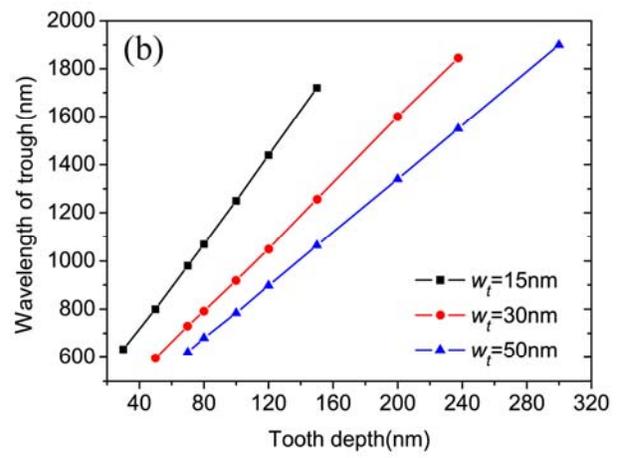

**Figure Captions**

Fig. 1. (Color online) Real part of the effective index of refraction versus the width of a MIM slit waveguide structure.

Fig. 2. (Color online) The structure schematics of a single tooth-shaped waveguide filter, with the slit width of $w$, the tooth width of $w_t$, and the tooth depth of $d$.

Fig. 3. (Color online) (a) Transmission of the single tooth-shaped MIM waveguide compared with a straight MIM slit waveguide. The width of the waveguide is $w$=50nm, and the tooth width and depth are respectively $w_t$=50nm and $d$=100nm. The contour profiles of field $H_y$ of the tooth-shaped waveguide at different wavelengths of (b) $\lambda$=510nm, (c) $\lambda$=783nm, and (d) $\lambda$=1550nm. In the FDTD simulation, we used the tabulation of the optical constants of silver given in Ref. 22.

Fig. 4. (Color online) (a) Transmission spectra of the waveguide filters with various tooth widths of $w_t$, at a fixed tooth depth of $d$=100nm and the slit width of $w$=50nm. (b) The wavelength of the trough vs. the tooth width of $w_t$.

Fig. 5. (Color online) (a) Transmission spectra of the waveguide filters with different tooth depths of $d$, and with a given tooth width of $w_t$=50nm and the slit width of $w$=50nm. (b) The wavelength of the trough vs. the tooth depth of $d$ with $w_t$=15nm, $w_t$=30nm and $w_t$=50nm.